\documentclass[aps, reprint, showpacs,superscriptaddress,floatfix,amsmath,amssymb]{revtex4-1}
\usepackage[cm]{fullpage}
\usepackage{float}
\usepackage{graphicx}
\usepackage{bm}
\begin{document}
\newcommand{\sub}[1]{_{\mbox{\scriptsize {#1}}}}

\preprint{AIP/123-QED}

\title[]{Homogeneous SPC/E water nucleation in large molecular dynamics simulations}
\author{Raymond Ang\'elil}
  \affiliation{Institute for Computational Science, University of Zurich, 8057 Zurich, Switzerland}
\author{J\"urg Diemand}
  \affiliation{Institute for Computational Science, University of Zurich, 8057 Zurich, Switzerland}
\author{Kyoko K. Tanaka}
  \affiliation{Institute of Low Temperature Science, Hokkaido University, Sapporo 060-0819, Japan}
\author{Hidekazu Tanaka}
  \affiliation{Institute of Low Temperature Science, Hokkaido University, Sapporo 060-0819, Japan}

\date{\today}
\begin{abstract}
We perform direct large molecular dynamics simulations of homogeneous SPC/E water nucleation, using up to $\sim 4\cdot 10^6$ molecules. Our large system sizes allow us to measure extremely low and accurate nucleation rates, down to $\sim 10^{19}\,\textrm{cm}^{-3}\textrm{s}^{-1}$, helping close the gap between experimentally measured rates $\sim 10^{17}\,\textrm{cm}^{-3}\textrm{s}^{-1}$. We are also able to precisely measure size distributions, sticking efficiencies, cluster temperatures, and cluster internal densities. We introduce a new functional form to implement the Yasuoka-Matsumoto nucleation rate measurement technique (threshold method). Comparison to nucleation models shows that classical nucleation theory over-estimates nucleation rates by a few orders of magnitude. The semi-phenomenological nucleation model does better, under-predicting rates by at worst, a factor of 24. Unlike what has been observed in Lennard-Jones simulations, post-critical clusters have temperatures consistent with the run average temperature. Also, we observe that post-critical clusters have densities very slightly higher, $\sim 5\%$, than bulk liquid. We re-calibrate a Hale-type $J$ vs. $S$ scaling relation using both experimental and simulation data, finding remarkable consistency in over $30$ orders of magnitude in the nucleation rate range, and $180\,$K in the temperature range. 
\end{abstract}

                            

\maketitle

\section{\label{sec:level1}Introduction}
The vapor-to-liquid transition of water is a common phenomenon in nature, relevant to many areas of technology and science. Attempts to predict the rate of homogeneous water nucleation often fail because of the lack of understanding of the properties of the tiny seeds of the intermediate phase, which are not necessarily large enough to have reached the bulk liquid properties. The relevant properties of the tiny clusters which affect predicted nucleation rates include surface tension, temperature, and density. Molecular dynamics simulation has proven to be a powerful test of thermodynamic analytical nucleation models, now that codes are efficient enough, and computers fast enough. Realistic, atmospheric nucleation rates are too low to be possible in direct computer simulations, due to the large number of molecules required. The lowest water nucleation rates performed in simulations and reported in the literature are $\sim 10^{23-24}\,\textrm{cm}^{-3}\textrm{s}^{-1}$\citep{tanaka2014,Tip4P_MD}, usually beyond the spinodal limit. Laboratory water nucleation rates on the other hand are far lower - usually $<10^{10}\,\textrm{cm}^{-3}\textrm{s}^{-1}$, although a few experiments have managed to measure far higher rates $\sim 10^{17}\,\textrm{cm}^{-3}\textrm{s}^{-1}$\citep{water_experiment10, water_experiment11,theory_comparison7}. Our simulations of homogeneous SPC/E water nucleation, which we report on in this paper, manage to close the gap considerably, resolving nucleation rates down to $\sim 10^{19}\,\textrm{cm}^{-3}\textrm{s}^{-1}$.

\begin{figure}[]
\includegraphics[scale = 0.155, trim={3cm 0 3.5cm 0},clip]{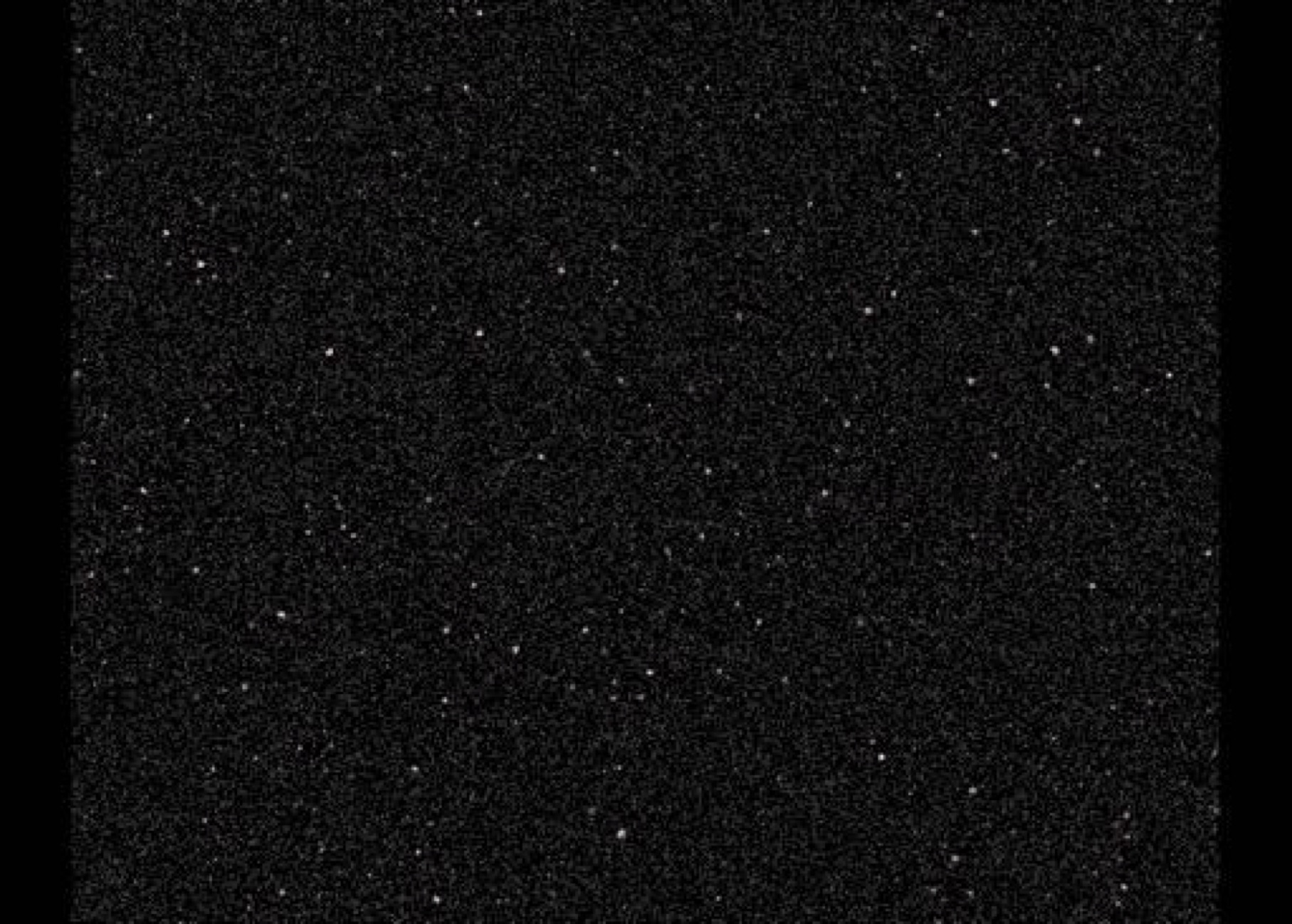}
\caption{A slice through the simulation T325f after 581$\,$ns. The color-map indicates the density, meaning that the white spots represent large clusters. By the end of the simulation, the largest cluster in this run has 527 members. This simulation box is $\sim 10\mu \textrm{m}\times 10\mu \textrm{m}\times 10\mu \textrm{m}$, although only a thin slice into the $z-$direction is visible.}\label{fig:whole_box}
\end{figure}

Nucleation models, which seek to provide explanations and predictions for nucleation rates, have a long history of falling short when compared to experimental results\citep{water_experiment1, water_experiment2, water_experiment3, water_experiment4, water_experiment5, water_experiment6,water_experiment7, water_experiment8, water_experiment9, water_experiment10, water_experiment11}. For the case of water, rate predictions from the classical nucleation theory disagree with experimental measurements by factors of $10^1-10^3$\citep{theory_comparison1,theory_comparison2,theory_comparison3, theory_comparison4, theory_comparison5, water_experiment2, theory_comparison7, theory_comparison8, theory_comparison9, theory_comparison10, water_experiment1, theory_comparison12}. These models also have difficulty when predicting rates measured in numerical molecular dynamics nucleation simulation experiments. However, with molecular simulation, one can make measurements more detailed and accurate than what's possible in laboratory experiments. Size distributions, nucleation rates, cluster densities, temperatures, and even cluster pressures, shapes, angular momenta, and surface tension measurements are possible. Understanding the properties of the tiny yet complex, many-body clusters which form is vital for the development of a complete and successful thermodynamic description of the phase transformation\citep{schweizer}.  Simulations allow us to identify the shortcomings in the assumptions made by existing nucleation models, and suggest ways they may be improved. Cluster properties are noisy, necessitating large systems with many millions of molecules. This demands costly compute power, and only recently have some of these direct measurement techniques become possible\citep{horsch,mdlj4,horsch2}.

Direct vapor-to-liquid molecular dynamics simulation for a Lennard-Jones fluid has become a popular exercise due the computational accessibility of the short-range, single-site potential\citep{mdlj1,mdlj2,mdlj3,mdlj4,mdlj5,mdlj6}. Water is significantly more demanding. For the same system size, more complicated molecular interaction potentials like SPC/E\citep{SPCE,mdwater1, mdwater2} and TIP4P\citep{tip4p_original,tip4p2005,Tip4P_MD} necessitate a few orders of magnitude more computational power than a pure Lennard-Jones simulation. An exception is mW water, a comparatively simple monoatomic single-site water model\citep{mw1,mw2,mw3,mw5,mw6,mw7,mwicenucleation}. MD nucleation simulations of mW water have been carried out, yet only on small systems with relatively high nucleation rates\citep{mwnucleation, mwicenucleation}. The monoatomic water model proposed by Zipoli et al. (2013)\citep{mdzipoli} offers similar advantages. However, we found that short-range potentials require extremely long equilibration times to form the correct equilibrium abundance of small clusters (dimers, trimers, etc.) in a supersaturated vapor, because interactions are rare, especially the three body encounters required for dimer formation. This drawback makes it computationally expensive to simulate realistic, steady state vapor-to-liquid with such short-range potentials - despite their low cost per time-step - and we do not use them in this work. Matsubara et al. (2007) \citep{mdwater1} simulate homogeneous vapor-to-liquid nucleation using the SPC/E water model and include a Lennard-Jones carrier gas, measuring rates down to $2.3\cdot 10^{25}\,\textrm{cm}^{-3}\textrm{s}^{-1}$. SPC/E simulations by Tanaka et al. (2014)\citep{tanaka2014} manage to reach nucleation rates $3\cdot 10^{24}\,\textrm{cm}^{-3}\textrm{s}^{-1}$. Both efforts additionally measure critical cluster sizes, formation energies, size distributions and sticking probabilities for systems in the $T=300-390\,$K, providing ample opportunity for model comparison and development.

In this study, we continue in similar spirit, yet simulating the SPC/E water vapor-to-liquid phase change in even larger computational volumes using longer time integrations. This allows for the measurement of lower nucleation rates than previously possible by a few orders of magnitude, and for the first time, measurements of naturally-formed SPC/E cluster density and temperature profiles. Our results provide opportunities for the verification and calibration of the standard assumptions which go into nucleation models, in a previously unexplored temperature and saturation regime.

\section{Simulations}
\subsection{Simulation code, setup and parameters}

We use the molecular dynamics SPC/E\citep{SPCE} water model. SPC/E is a rigid 3-site model, which registers Coulombic interactions, as well as polarization corrections to each site, and further adds a Lennard-Jones component to the oxygen atom potential. 

The Large-scale Atomic/Molecular Massively Parallel Simulator (or LAMMPS) computer program\citep{lammps}, developed at the Sandia National Laboratories and distributed under the GPL license, was used to perform the SPC/E simulations. We have verified that our runs produce the same results as found in similar, yet smaller SPC/E numerical nucleation experiments\citep{tanaka2014}, which used an independent molecular dynamics code. We cut the short range Lennard-Jones component to the force field off at $9.8\,$\AA. For these forces, as well as the others, the interactions are computed directly on per atom. However, the SPC/E Coulombic interactions are long range, and so after $63\,$\AA, the spectral solver takes over, and the interactions computed in reciprocal space. LAMMPS uses a particle-particle/particle-mesh solver. The solver maps the atom charges onto a mesh, solves the Poisson equation (Maxwell's equation for the electric field) by performing a 3D fast Fourier transform, then interpolates the electric fields on the mesh points back onto the atom positions\citep{lammps,hockneycomputer,Pollock1996}. SPC/E molecule rigidity is ensured through the use of the SHAKE algorithm\citep{shake}. We choose an integration time-step of  $\Delta t=2\,$fs, common for SPC/E water simulations\citep{tanaka2014}. A typical simulation runs for 72 hours on 1024 cores. Our largest simulation ran for 1000 hours on 8192 cores on the Piz Daint supercomputer at Centro Svizzero di Calculo Scientifico (CSCS), performing $3\cdot 10^8$ integration time-steps. 

The simulation box has periodic boundary conditions. Initially the molecules are given random non-overlapping positions and random velocities. This is done at $1000\,$K, after which the ensemble is cooled and the box size expanded until the simulation reaches the target temperature and pressure. The run continues in this state under NVT conditions, regulated by a Nose-Hoover thermostat\citep{nosehoover1,nosehoover2,shioda} with temperature damping timescales of $1000\,$fs. 

At this stage the gas is allowed to equilibrate for a fixed amount of time - dependent on the run temperature (Refer to table \ref{tab:thermo} for the chosen equilibration timescales $t_e$ at each temperature). During this phase the subcritical cluster equilibrium distribution forms. Around this stage we begin to make nucleation rate, size distribution, and cluster growth rate measurements. For most runs, the nucleation rate is low enough that unnatural effects from the interventions due to the thermostat are minimal. Our nucleation rates are low enough that the latent heat of transformation in the simulations is extremely small, resulting in only a faint influence from the thermostat. Our largest run sees a total energy increase of $\sim 0.1\%$ over the steady-state phase, i.e. our simulations are very close to NVE (micro-canonical) ensembles.

The first few columns of table \ref{tab:results} lists the runs which were carried out, their target temperatures, box sizes, number of molecules, and their run times.

\begin{table}[t]
\caption{Thermophysical quantities and parameters at each temperature. The vapor equilibrium pressure $P_v$, the planar surface tension $\gamma$, and the  bulk liquid density $\rho_b$ for SPC/E water are determined from the fitting functions in Matsubara et al. (2007)\citep{mdwater1}. $\eta$ and $\xi$ are nucleation model parameters\citep{mdlj3}. $t_{\textrm{e}}$ the time over which we allow our simulations to equilibrate into the steady-state before taking size distribution measurements.}
\begin{tabular}{ c c c c c c c c}
\hline  \hline
T  		& $P_{\textrm{v}}$ 	& $\gamma$      &$\rho_b$       & $r_0$        			&$\eta$	&   $\xi$	& $t_{\textrm{e}}$	\\ 
$[\textrm{K}]$ 	& [dyn/cm$^2$] 			& [dyn/cm]		&[g/cm$^3$]	 	& [$10^{-8}\,$cm] 		  			&		&			&	[ns]\\ \hline
300 	&  $8.9\cdot10^3$		& 53.4			& 0.997	 		& $1.93 $   & 6.05 	& 8.95 		& 25\\ 
325 	&  $4.1\cdot10^4$ 		& 50.1			& 0.982	 		& $1.94 $   & 5.29 	& 7.47 		& 20\\ 
350 	&  $1.49\cdot10^5$ 		& 46.6			& 0.966	 		& $1.95 $   & 4.63 	& 6.31 		& 10\\ 
375 	&  $4.47\cdot10^5$ 		& 42.9			& 0.946	 		& $1.97 $   & 4.02 	& 5.38 		& 6\\ 
\hline
\end{tabular}\label{tab:thermo}
\end{table}

\begin{table*}[t]
\caption{Run temperature $T$, supersaturation $S$ as calculated from the run monomer number density, box length $L$, molecule number $N$, runtime $t_{\textrm{end}}$, nucleation rate measured from simulation $J_{\textrm{MD}}$, critical cluster size $i^*$ from the first nucleation theorem, $J_{\textrm{SP}}$ Semi-phenomenological model prediction, $J_{\textrm{MCNT}}$ modified classical nucleation theory prediction, $i_{\Delta G}^*$ critical size from the $\Delta G$ reconstruction. }
\begin{tabular}{ c | c c c c c| c c c c c c }
\hline \hline
 Run ID 	& $T$           &$S$ &$L$    	&$N$			&   $t_{\textrm{end}}$               &$J_{\textrm{MD}}$						&   $i^*$		& $\alpha$ &$J_{\textrm{SP}}$ &$J_{\textrm{MCNT}}$ &$i_{\Delta G}^*$\\  
            & $[\textrm{K}]$&					&$[\textrm{nm}]$ & $[\cdot 10^3]$  & $[\textrm{ns}]$&$[\textrm{cm}^{-3}\textrm{s}^{-1}]$&				&		   &$[\textrm{cm}^{-3}\textrm{s}^{-1}]$&$[\textrm{cm}^{-3}\textrm{s}^{-1}]$&\\ \hline 
 T300a 		&$300$			&$23.30 \pm 2.89$&$4859.5$		& $768$ 		&   $31.5$	&$1.56 \pm 0.66\cdot 10^{24}$ 	& - 			& $1.31 $ & $4.42\cdot 10^{24}$& $5.68\cdot 10^{26}$&9\\ 
 T300b  	&$300$			&$19.61 \pm 1.71$&$6581.2$		& $1500$ 	& 	$43$	&$3.29 \pm 1.95\cdot 10^{23}$ 	& - 			& $1.11 $ & $1.25\cdot 10^{24}$& $3.13\cdot 10^{26}$&9\\ 
 T300c  	&$300$			&$13.44 \pm 0.94$&$7591.9$		& $1500$		&   $51$ 	&$5.93 \pm 2.04\cdot 10^{22}$ 	& - 			& $0.59 $ & $4.26\cdot 10^{22}$& $6.49\cdot 10^{25}$&11\\ \hline
 T325a      &$325$			&$7.54 \pm 0.46$&$3307.0$		& $324$ 		& $55.0$ 	&$3.89 \pm 1.14\cdot 10^{23}$ 	& - 			& $0.52 $ & $1.00\cdot 10^{23}$& $1.80\cdot 10^{26}$&11\\
 T325b     	&$325$			&$6.70 \pm 0.33$&$3441.6$		& $324$ 		& $82.5 $ 	&$1.90 \pm 0.15\cdot 10^{23}$ 	& $11\pm7$ 		& $0.59 $ & $1.93\cdot 10^{22}$& $8.02\cdot 10^{25}$&13\\
 T325c		&$325$ 			&$6.03 \pm 0.20$&$4863.0$		& $768$ 		& $113.6 $	&$4.42 \pm 0.21\cdot 10^{22}$ 	& $17\pm6$		& $0.26 $ & $3.60\cdot 10^{21}$& $3.47\cdot 10^{25}$&13\\
 T325d		&$325$ 			&$5.20 \pm 0.06$&$6461.7$		& $1500$ 	& $177.0 $		&$2.77 \pm 0.47\cdot 10^{21}$ 	& $20\pm5$		& $0.25 $ & $2.23\cdot 10^{20}$& $8.35\cdot 10^{24}$&17\\
 T325e		&$325$ 			&$4.59 \pm 0.03$&$8167.4$		& $2592$ 	& $228.6$	&$2.24 \pm 1.18\cdot 10^{20}$ 	& $21\pm3$		& $0.15 $ & $1.30\cdot 10^{19}$& $1.86\cdot 10^{24}$&24\\
 T325f		&$325$ 			&$4.15 \pm 0.0003$&$9895.9$		& $4116$ 	& $581.2$	&$1.80 \pm 0.36\cdot 10^{19}$ 	& $23\pm4$		& $0.16 $ & $8.35\cdot 10^{17}$& $4.21\cdot 10^{23}$&-\\  \hline
 T350a		&$350$ 			&$4.60 \pm 0.15$&$2573.7$		& $324$ 		& $38.8$	&$7.15 \pm 0.83\cdot 10^{23}$ 	& -				& $0.38 $ & $8.16\cdot 10^{22}$& $2.50\cdot 10^{26}$&13\\ 
 T350b		&$350$ 			&$4.29 \pm 0.09$&$2680.1$		& $324$ 		& $34.0$	&$1.31 \pm 0.44\cdot 10^{23}$ 	& $22\pm7$		& $0.23 $ & $2.19\cdot 10^{22}$& $1.26\cdot 10^{26}$&16\\
 T350c		&$350$ 			&$3.91 \pm 0.001$&$2783.2$		& $324$ 		& $75.6$	&$2.10 \pm 0.48\cdot 10^{22}$ 	& $19\pm5$		& $0.21 $ & $2.97\cdot 10^{21}$& $4.35\cdot 10^{25}$&18\\
 T350d		&$350$ 			&$3.58 \pm 0.003$&$2893.7$		& $324$ 		& $161.8$	&$3.85 \pm 0.61\cdot 10^{21}$ 	& $27\pm2$		& $0.18 $ & $3.20\cdot 10^{20}$& $1.29\cdot 10^{25}$&21\\
 T350e		&$350$ 			&$3.31 \pm 0.001$&$4993.6$		& $1500$ 	& $232.6$	&$2.02 \pm 0.15\cdot 10^{20}$ 	& $37\pm4$		& $0.06 $ & $3.15\cdot 10^{19}$& $3.54\cdot 10^{24}$&-\\  \hline
 T375a		&$375$ 			&$3.2 \pm 0.06$	&$2017.4$		& $324$ 		& $18.0$	&$1.43 \pm 0.13\cdot 10^{24}$ 	& $30\pm10$		& $0.40 $ & $7.31\cdot 10^{22}$				   & $3.14\cdot 10^{26} $               &18 \\
 T375b		&$375$ 			&$2.97 \pm 0.005$&$2107.5$		& $324$ 		& $38.2$	&$1.60 \pm 0.12\cdot 10^{23}$ 	& $30\pm6$		& $0.26 $ & $1.03\cdot 10^{22}$& $1.06\cdot 10^{26}$&20\\
 T375c		&$375$ 			&$2.83 \pm 0.001$&$2158.6$		& $768$ 		& $52.4$	&$4.64 \pm 1.26\cdot 10^{22}$ 	& $35\pm3$		& $0.23 $ & $2.42\cdot 10^{21}$& $4.69\cdot 10^{25}$&21\\
 T375d		&$375$ 			&$2.73 \pm 0.001$&$2937.3$		& $1500$ 		& $55.4$	&$7.22 \pm 1.13\cdot 10^{21}$ 	& $37\pm6$		& $0.25 $ & $7.34\cdot 10^{20}$& $2.36\cdot 10^{25}$&-\\  \hline
\end{tabular}\label{tab:results}
\end{table*}

\subsection{Simulation Analysis}\label{sec:analysis}
We use the simple Stillinger criterion\citep{stillinger} (also known as the friends-of-friends method) to identify clusters. As the simulation runs, the cluster size distribution is regularly calculated and outputted, typically resulting in $>1000$ size distribution histograms per simulation. The linking length was set at $6\textrm{\AA}$ for all runs, and was tested to yield stable size distributions under convergence tests. Furthermore, this choice yields a monomer-dimer number ratio consistent with what is expected from the second virial coefficient applied to the SPC/E interaction potential\citep{mdwater1}. The regularly-outputted size distributions can then be converted into cluster threshold sizes, whose slopes in the steady-state regime are the nucleation rates. Refer to section \ref{sec:jrates} for further details on the nucleation rate analysis, as well as the results. From the nucleation rate vs. supersaturation ratio landscape, we calculate the critical cluster sizes using the first nucleation theorem\citep{CNT5,SP1}.The size distributions also allow us to follow the growth rate of the largest clusters in each simulation, providing a measurement of the monomer-cluster interaction sticking efficiency (refer to section \ref{sec:alpha}). 

The measurements of specific cluster properties and how they vary with cluster size is crucial to testing assumptions used in theoretical nucleation models. However, because of the noisy nature of many of these properties, one needs many millions of molecules per simulation in order to resolve interesting cluster properties. Due to this limitation, we perform cluster temperature and cluster density profile measurements only for our largest simulation, which contained $\sim 4\times 10^6$ molecules. We perform this at the end of the simulation, well-within the steady-state nucleation regime. This calls for per-atom outputs of velocity and position information. Sections \ref{sec:densities} and \ref{sec:temps} detail how the density profile and temperature measurements respectively are made, and discuss the results.

\section{Nucleation Rates}\label{sec:jrates}

\begin{figure*}[]
\includegraphics[scale = 0.55]{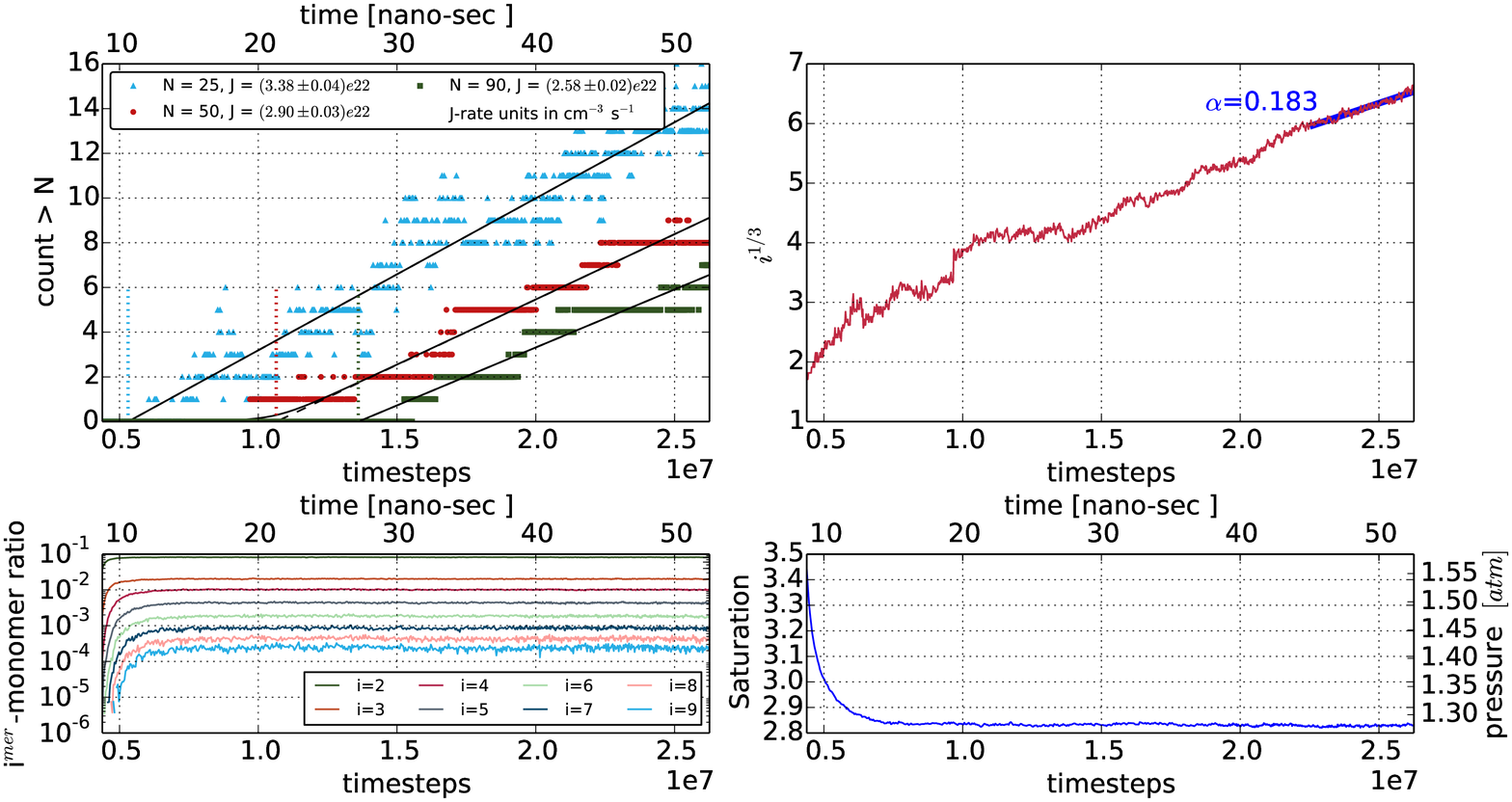}
\caption{For the $768000$ molecule simulation T375c, clockwise from the upper left panel: (1) Nucleation rate measurements, (2) largest-cluster growth curve, (3) number density and monomer partial pressure, and (4) $i^{\textrm{-mer}}$ concentrations, all over the entire run period. Nucleation rates are measured by counting the number of clusters above a specified threshold size, at periodic time intervals. The steady-state regime slope is the nucleation rate. (Refer to section \ref{sec:analysis} and equation \eqref{eq:nucleation_rate}.) The dotted vertical lines indicate the lag times for each size. The cluster sticking probabilities $\alpha$ are calculated from the measured slopes, $di/dt$, using equation \eqref{eq:alpha}. We can see that this run took $\sim 20\,$ns to equilibrate, and spent another $\sim 5-15\,$ns in the lag phase before reaching the steady-state regime (for the chosen threshold sizes of $N=50$ and $N=90$). The probability that a cluster-monomer encounter results in the cluster growing by one molecule is the sticking probability $\alpha = 0.22.$}\label{fig:spce_t375_r2-25-50} 
\end{figure*}

%\begin{figure*}[]
%\includegraphics[scale = 0.55]{spce_t325_r8.pdf}
%\caption{For the $4'116'000$ molecule simulation t325f, clockwise from the upper left panel: (1) Nucleation rate measurements, (2) largest-cluster growth curve, (3) number density and monomer partial pressure, and (4) $i^{\textrm{-mer}}$ concentrations, all over the entire run period. This run took $\sim 80\,$ns to equilibrate, and spent another $\sim 40\,$ns in the lag phase before reaching the steady-state regime (for the chosen threshold sizes $N=50$). The probability that a cluster-monomer encounter results in the cluster growing by one molecule is $\alpha = 0.175$. Note the size distribution trimer anomaly (2nd line in the lower left panel), also observed in previous SPC/E simulations\citep{tanaka2014}: at lower temperatures there is a pronounced deficit in the number of trimers, relative to the number of dimers and tetramers. We do not expect that this anomaly affects the nucleation rate, as the critical size for this run is far larger, at $\sim 22$. }\label{fig:spce_t325_r8}
%\end{figure*}

We use a modified Yasuoka-Matsumoto method\citep{mdwater2} (threshold method) to measure nucleation rates. In the steady-state nucleation regime, the time rate of increase of the number of clusters above a certain size $N$ is the nucleation rate. However, the simulations must equilibrate - form the sub-critical size distribution - and properly populate it before they reach the steady state nucleation regime. How long the simulations take to transition into the steady state regime is not known \textit{a priori}. Nucleation rates estimated from the first nucleation event alone (e.g. mean first passage time or survival probability methods) can be orders of magnitude smaller than the true steady state nucleation rates\citep{shneidman,mokshin}. Thus we use the following method: To the size-threshold curves, we fit the following function, which is able to capture the transition from the equilibration to the steady-state phase, 
\begin{equation}\label{eq:nucleation_rate}
N(>i) = J \cdot \mathcal{N}\left(>i, t\right) + \mathcal{N}\left(>i, 0\right),
\end{equation}
where
\begin{equation}
\mathcal{N} \left(>i, t\right) = \left[\frac{\pi}{2} + \arctan{\frac{t-t_0}{t_{\textrm{r} }}} \right] \cdot \frac{t-t_0}{\pi} \cdot V ,
\end{equation}
where $J$ is the nucleation rate, $t_0$ is the lag time, $t_r$ is the relaxation timescale and $V$ the volume of the simulation box. This function captures the system's transition from the initial equilibration phase to the intermediate relaxation phase as the clusters begin to form, through to the steady-state regime. We count clusters above a certain post-critical size $N$, frequently throughout the simulation, and fit the count to this curve, allowing $J$, $t_0$, and $t_{\textrm{r} }$ to vary. Visual inspection of this approach is provided in the upper left panels of figures \ref{fig:spce_t375_r2-25-50} for run T375c respectively.

Our nucleation rate measurements are listed in \ref{tab:results}. Figure \ref{fig:j-rate_overview} plots our simulations' nucleation rates against supersaturation, and includes comparison to earlier results\citep{tanaka2014}, which used smaller simulations and were therefore restricted to lower nucleation rates. Estimates for the critical cluster sizes using the first nucleation theorem, via
\begin{equation}
i^*_{\textrm{NT}} = \left(\frac{\partial \ln J}{\partial \ln S} \right)_\textrm{T} - 1
\end{equation}
 are included as annotations. Our nucleation rate results can be split into two categories:

\begin{itemize}
\item \textbf{High temperature} ($325\,\textrm{K}$, $350\,\textrm{K}$, $375\,\textrm{K}$): Runs at these temperatures have nucleation rates in the range $\sim 10^{19-24}\,\textrm{cm}^{-3}\textrm{s}^{-1}$. These runs have generally low errors on the nucleation rates. For the higher nucleation rates, there is an error on the supersaturation, as the pressure drops significantly due to the large number of clusters forming quickly.  
\item \textbf{Low temperature} ($300\,\textrm{K}$): Here we measure nucleation rates in the range $\sim 10^{23-24}\,\textrm{cm}^{-3}\textrm{s}^{-1}$. These runs suffer from extremely long equilibration periods, which continue while the initial large, stable clusters are already forming. In other words, the sub-critical distribution formation timescale $t_r$ is longer than the nucleation timescale $1/(J\cdot V)$. This leads to large errors in both the nucleation rate measurements and the supersaturation measurements. 
\end{itemize}

\section{Rate comparison with analytical models}
\begin{figure}[]
\includegraphics[scale = 0.4]{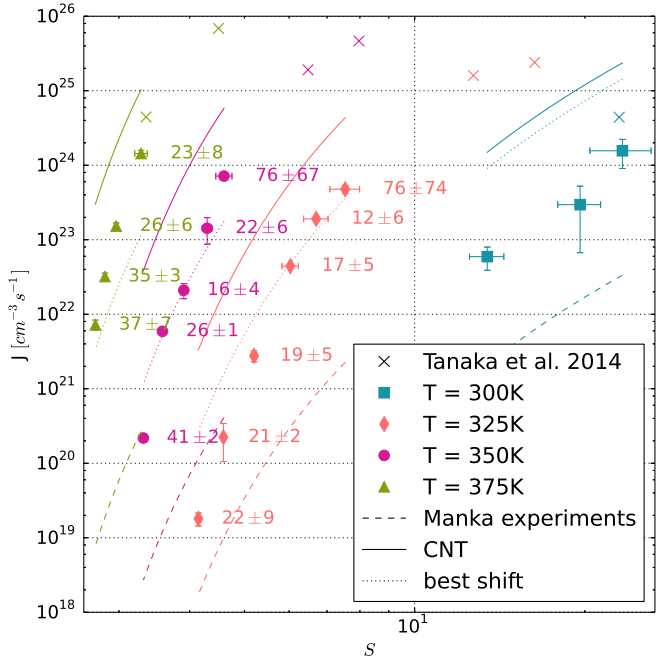}
\caption{The filled solid markers are the nucleation rates we measure from our simulations. The solid curves correspond to the classical nucleation theory \eqref{eq:cnt} predictions. The dashed curve includes the CNT correction factor \eqref{eq:cnt_corr} used in Manka et al. (2011)\citep{water_experiment1}, and the dotted one is our best-fit correction factor (the fit excludes the runs at $T=300\,$K).}\label{fig:j-rate_overview}
\end{figure}

Nucleation models endeavor to describe the phase change process in purely thermodynamic terms. The standard approach tries to find the balance between the Gibbs free energy gain and cost due to the creation of volume and surface. The classical nucleation theory (CNT)\citep{CNT1,CNT2,CNT3,CNT4,CNT5,CNT6,CNT7} is the most basic of them all, and forms the basis upon which many appendages have since been added. In the CNT, the surface energy term in the Gibbs free energy is simply calculated using the planar surface tension, with no additional corrections. 
The CNT nucleation rate is \citep{beckerCNT}
\begin{equation}\label{eq:cnt}
J_{\textrm{CNT}} = \sqrt{\frac{32\pi\gamma}{9 m}}r_0^3 \left(\frac{p_{\textrm{g}}}{k_b T} \right)^2 \exp\left[\frac{256\pi^2 r_0^3\gamma^3}{27\left(k_B T\right)^3\left(\log S \right)^2} \right],
\end{equation}
where $m$ is the molecular mass, $\gamma$ the planar surface tension at the run temperature, $p_{\textrm{g}}$ the monomer partial pressure
in the simulation box (assuming an ideal gas) gas pressure and $S$ the supersaturation
\begin{equation}
S = \frac{p_\textrm{g}}{p_{\textrm{v}}},
\end{equation}
where $p_v$ is the equilibrium vapor pressure at the run temperature. $r_0$ is a characteristic molecular radius:
\begin{equation}
r_0 = \left(\frac{3}{4\rho_l \pi} \right)^{1/3},
\end{equation}
where $\rho_l$ is the bulk liquid density at the run temperature. Table \ref{tab:thermo} includes the thermodynamic variables for SPC/E water, which we use in our analysis and comparison to nucleation models. The CNT predictions for the nucleation rates of SPC/E water at our runs' supersaturations are shown as solid curves in figure \ref{fig:j-rate_overview}. We find that the CNT predicts too-high nucleation rates by factors of $10^{1-2}$.

Various authors \citep{correction_formula,water_experiment3,water_experiment1} employ a 2-parameter, temperature dependent correction factor
\begin{equation}\label{eq:cnt_corr}
J_{\textrm{corr}} = J_{\textrm{CNT}}\exp\left(A +\frac{B}{T} \right).
\end{equation}
The Manka et al. (2010)\citep{water_experiment1} (see their figure 4) laminar flow diffusion chamber experiments find that the parameter pair $(A,B) = (-27.56, 6500\,$K$)$ corresponds to a global fit of their results and previous experiments\citep{water_experiment2,water_experiment3,water_experiment4,corrmore,water_experiment6,water_experiment7,water_experiment8,water_experiment9,theory_comparison12,corrmore2}. 
With these parameters the CNT rate prediction remains unchanged at a temperature of $-B/A = 235.8\,$K, and they still increase with temperature (at a fixed $S$), but less strongly than in CNT.
These corrected CNT predictions, when extrapolated to our supersaturations (dashed curves in figure \ref{fig:j-rate_overview}) under-predict our measurements by 1-3 orders of magnitude. Using our data at temperatures $T=325,350,375\,$K to determine the best-fit parameter pair, we find $(A,B) = (-20.5 ,6100)\,$K. With our parameter pair the CNT rate prediction remains unchanged at a temperature of $-B/A = 297.6\,$K, and they increase with temperature at an rate between CNT and the Manka et al. model. However, we note that the resulting curves (dotted lines in figure \ref{fig:j-rate_overview}) are not quite steep enough - casting doubt on whether a purely temperature-dependent correction is sufficient in this high supersaturation regime.  

\begin{figure}[]
\includegraphics[scale = 0.53]{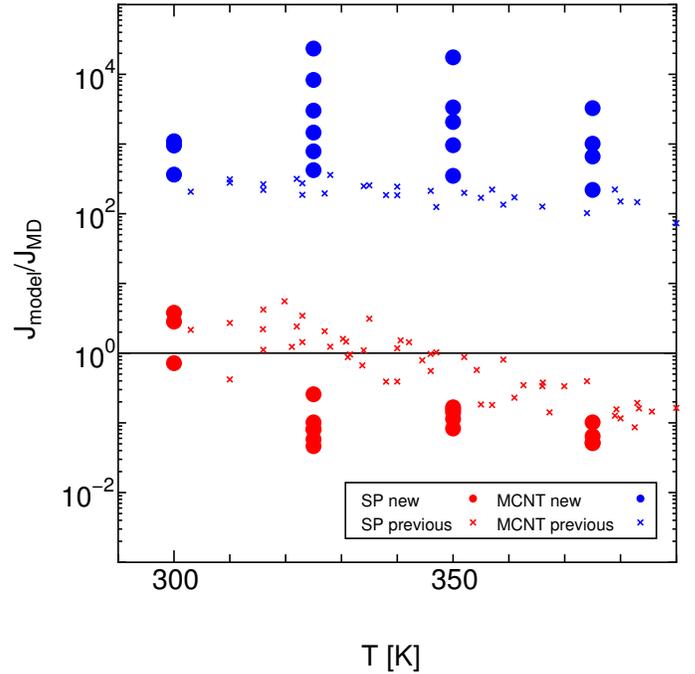}\caption{A comparison between the measured molecular dynamics nucleation rates, and those predicted by the analytic SP and MCNT nucleation models. `new' and `previous' refer to the simulations detailed in this paper, and those by Tanaka et al. (2014)\citep{tanaka2014} respectively. While the SP model is more reliable than the MCNT, the SP model shows a trend of rate under-prediction for lower supersaturations.}\label{fig:compare-j}
\end{figure}

The Modified Classical Nucleation Theory (MCNT)\citep{mdlj2} implements a minor modification to the CNT, namely, it stipulates that the free energy of formation of a cluster of size one is zero. This results in a free energy shift for all cluster sizes. Like the CNT, the MCNT over-predicts the nucleation rates and here the differences are even slightly larger (factor of $10^{2-4}$). Figure \ref{fig:compare-j} shows the ratio between the MCNT model predictions (red markers) and the direct MD measurements. Refer to table \ref{tab:results} for the MCNT model nucleation rate predictions. 
 
The Semi-Phenomenological model (SP)\citep{SP1,SP2,SP3,SP4,SP5} attaches a $\sim 1/R$ (or $\sim i^{-1/3}$) correction to the surface tension, where $R$ is the cluster size under the assumption of sphericity. This radial dependence is functionally equivalent to that introduced by the Tolman length\citep{Tolman1949,water_experiment6,tolman_water}, although the motivation is different: the coefficient to this term is set by the second virial coefficient $B_2$\citep{mdwater1} so that the dimer number density is correctly predicted. The nucleation rate predictions for the SP model relative to the measured values are plotted with red markers in figure  \ref{fig:compare-j}. The predictions at $T=300\,$K are somewhat accurate - within a factor of $5$ of the measured values, although, as noted in section \ref{sec:jrates} the measurements at these temperatures carry significant uncertainty in the nucleation rate. At the higher temperatures, the SP model under-predicts the measured MD rates by factors of $4-80$. Table \ref{tab:results} lists the SP model nucleation rate predictions.

\section{Nucleation rate scaling}

In this section we examine the scaling of the nucleation rates.  For the case of water, Hale (2005)\citep{hale} (and similarly for Lennard-Jones in Hale (2010)\citep{moarhale})
uses a scaling relation\citep{superheighl} for experimentally measured nucleation rates over the range $J=10^{4-10}\,$cm$^{-3}$s$^{-1}$ of 

\begin{equation} \label{eq:scaling1}
\frac{\ln S}{(T_{\rm
  c}/T-1)^{1.5}}.
 \end{equation} 
 
\begin{figure}[]
  \includegraphics[scale =
  0.55]{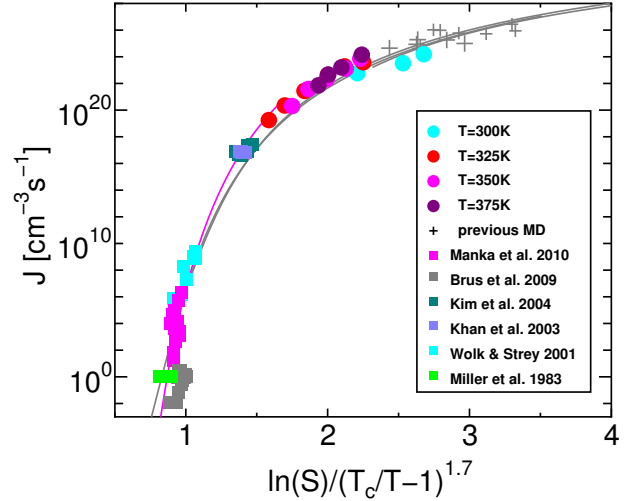}\caption{Nucleation
    rates as function of $\ln S /(T/T_{\rm c}-1)^{1.7}$ for MD
    simulations and experiments\citep{tanaka2014,water_experiment1,water_experiment9,water_experiment10,theory_comparison7,water_experiment3,water_experiment8}.  Our simulations are filled circles
    and the previous simulations `+' symbols.  The solid curves show
    the SP model for various temperatures ($210$, $315$, $350$, and $380\,$K).
    For thermodynamic quantities such as the surface tension and the
    saturated vapor pressure, we use those of the SPC/E at 315, 350,
    and $380\,$K for the comparison with the MD results, while real water
    at $210\,$K (pink curve) for the comparison with the experiment.
  }\label{fig:compare-jscale-new-largeMD-d2}
\end{figure}

Tanaka et al. (2014)\cite{mdlj5}
showed that this scaling relation works well for large scale Lennard-Jones simulations and Argon laboratory experiments,
albeit with an exponent of 1.3 instead of 1.5.
We confirm that the same scaling relation \eqref{eq:scaling1} applies well to our SPC/E water nucleation rate measurements. However, we find that 
the combined nucleation rates from both SPC/E
simulations and laboratory experiments with water are even better scaled by 
\begin{equation}\label{eq:scaling2}
\frac{\ln S}{(T_{\rm
  c}/T-1)^{1.7}}.
\end{equation}
 Figure~\ref{fig:compare-jscale-new-largeMD-d2} shows
the nucleation rates as a function of \eqref{eq:scaling2}.
This empirical scaling relation seems to work well over a surprisingly
wide nucleation rate range - from   
$J=10^{-2}$ to $J=10^{28}\,$cm$^{-3}$s$^{-1}$ for both MD simulations and
experiments. The results from the MD simulations join smoothly with
the experiments with the scaling by $\ln S /(T_{\rm c}/T-1)^{1.7}$.
Figure~\ref{fig:compare-jscale-new-largeMD-d2} also shows, using solid curves, the
nucleation rates predicted by the SP model for various temperatures.
This scaling relation also works very well for the SP model.
 
\section{Sticking Probabilities}\label{sec:alpha}
\begin{figure}[]
\includegraphics[scale = 0.65]{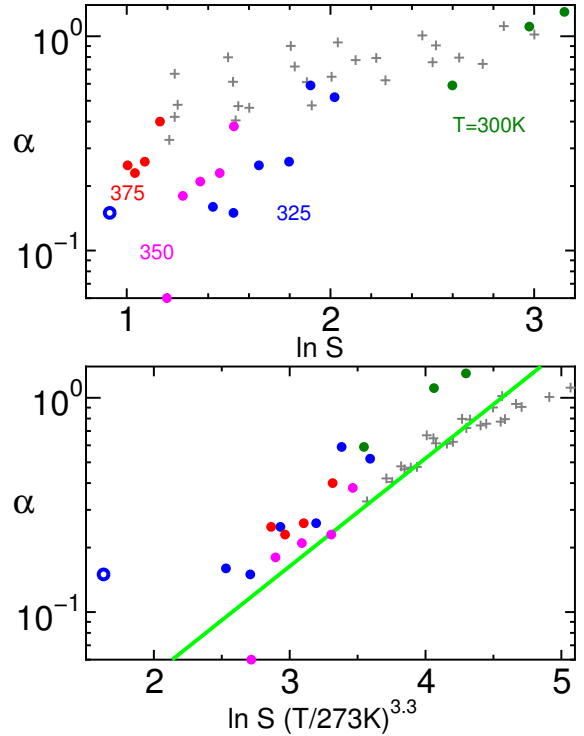}
\caption{Upper panel: sticking probability measurements for our simulations (solid markers) and from previous simulations (`+' symbols)\citep{tanaka2014}. Our simulations continue the expected trend of lower growth rates with decreasing gas pressure. The lower panel shows a temperature-dependent scaling relation which reduces the results to a single curve. The blue donut marker indicates the sticking probability measured under \textit{experimental} saturations (refer to the end of section \ref{sec:alpha}).   }\label{fig:alpha-largeMD}
\end{figure}

The sticking probabilities $\alpha$ can be calculated from the rate at which large, stable clusters grow. For each run, we observe the size of the largest cluster, and measure its growth rate $di/dt$ over the second half of the simulation. Early on in the simulations, before stable clusters have formed, the \textit{largest} designation jumps between clusters. However, the first stable cluster to form is likely to remain the largest until the end of the simulation. The upper right panel of figure \ref{fig:spce_t375_r2-25-50} shows our cluster growth rate measurements for run T375c. We find the cluster size $i\left(t\right)$ to be strongly cubic within the steady-state regime. The cluster growth rate is therefore proportional to the surface area. This is consistent with what has been found in Lennard-Jones nucleation simulations\citep{mdlj2,mdlj3}. We may determine $\alpha$ \citep{tanaka2014, mdlj3} from
\begin{equation}\label{eq:alpha}
\alpha = \frac{3}{4\pi r_0^2 v_{\textrm{th}}n\left(1\right) }\left(1 - \frac{1}{S} \right)^{-1} \frac{di^{1/3}}{dt}.
\end{equation}
The supersaturation $S$ dependence includes the effect of the evaporation of molecules from the clusters into the gas. We list the measured sticking probability results in table \ref{tab:results}. Sticking probability results for our low temperature $T=300\,$K runs are somewhat unreliable and can exceed unity due to the large number of dimers, trimers, and tetrames which also contribute to cluster growth. Eq \eqref{eq:alpha} considers the accretion and evaporation of monomers only. Our sticking probability measurements are consistent with those measured at slightly higher supersaturations in Tanaka et al. (2014) \citep{tanaka2014}. The upper panel of figure \ref{fig:alpha-largeMD} plots $\alpha$ against $S$. The sticking probability is a necessary prerequisite for performing the $\Delta G$ landscape reconstruction procedure for post-critical clusters (see section \ref{sec:delG}). 

While we are, due to computational constraints, unable to probe the low nucleation rates observed in laboratory experiments, it is possible to measure cluster growth rates under laboratory conditions. We have performed an additional simulation from the end state of T325c, in which we measured a sticking probability $\alpha = 0.26$. We target the temperature and saturation conditions found in Brus et al. (2008)\citep{theory_comparison10}. Using a Nose-Hoover thermostat we maintain the temperature, and gently increase the box size until the supersaturation $S= 2.5$, after which we continue running for $60\,$ns. Under these low pressure conditions, no new clusters nucleate (Brus et al. (2008)\citep{theory_comparison10} report nucleation rates $\sim 10^1\,\textrm{cm}^{-3}\textrm{s}^{-1}$) due to our comparatively small and short-lived system. However, clusters which had previously nucleated and then grown under the original conditions, persist. Using the largest of these still-post critical clusters, we measure a decreased growth rate: an $i^{1/3}$ slope shallower by a factor of $\sim7$. Including this, and the reduced (by a factor $\sim 3$) monomer number density into Eq. \eqref{eq:alpha} gives a sticking probability for these laboratory-like growth rates of $\alpha=0.15$. In nucleation models the sticking efficiency is usually taken to be unity, entering linearly in the transition growth rate (typically denoted $R^+$), as a prefactor to the $\Delta G_i$ exponent. We find that in the $T=325\,$K and $S=2.5$ regime, the water monomer-cluster sticking efficiency is approximately one seventh of what is usually used in model predictions, implying an expected lowering of predicted nucleation rates by the same factor. 

\section{Free energy reconstruction}\label{sec:delG}

\begin{figure}[]
\includegraphics[scale = 0.5]{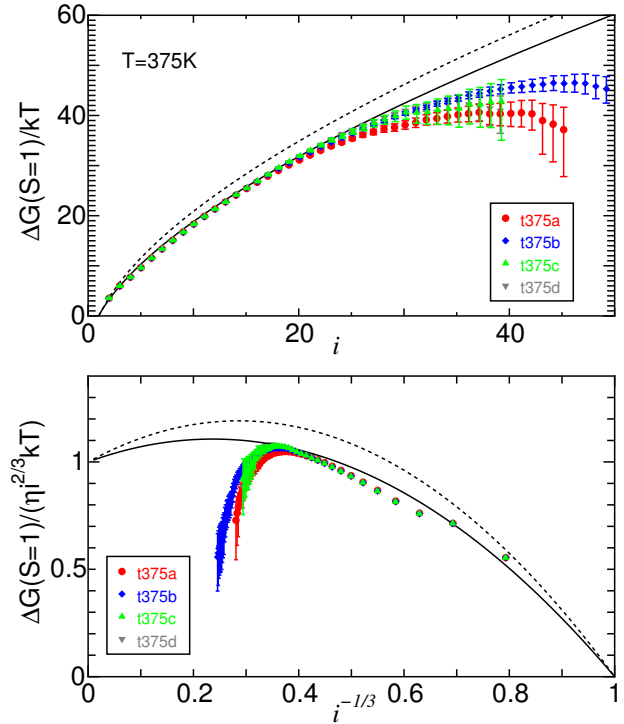}
\caption{Reconstructed Gibbs free energy curves shifted to $S=1$ (top panel). Runs at T = 375 K
at different supersaturations were used. To get to $S=1$ the CNT volume term was subtracted, the
resulting $\Delta G_i (S = 1)$ can be interpreted as the surface term. The bottom panel shows
the $\Delta G_i (S = 1)$ divided by the surface term from CNT. The surface term from the SP model
and a simple fitting function (Eq. \eqref{eq:kyoko}) are shown with dotted and solid lines.}\label{fig:delgs-t375}
\end{figure}

In this section, we evaluate the formation free energy of a cluster $\Delta G_i (S)$
directly from our molecular dynamics simulations, even for post-critical cluster sizes.  We obtain $\Delta G_i (S)$ from the
equilibrium size distribution of the cluster. The equilibrium size
distribution can be obtained using the steady state size distribution,
the accretion rate of molecule on a cluster, and the nucleation rate,
all of which can be measured directly in from the MD simulations.
Refer to Tanaka et al. (2014)\cite{mdlj5} for a thorough explanation of the technique.  The
cluster size distributions are measured in the MD simulations and
time-averaged over the steady state nucleation phase.  In the
accretion rate, we use the value of the sticking probability obtained
from MD simulations.  With the use of them, we reconstruct the 
full equilibrium size distribution (at all sizes $i$, where we have good abundance estimates, including $i >> i^*$) and then the entire
free energy function $\Delta G_i (S = 1)$. We can further derive $\Delta G_i (S = 1)$,
which is a surface term corresponding to the work required to form the
vapor-liquid interface, by subtracting the volume term from $ \Delta G_i (S)$:
\begin{equation}
\Delta G_i (S=1) =\Delta G_i (S) + (i-1) \ln S. 
\label{deltags1}
\end{equation}
Figure~\ref{fig:delgs-t375} shows $\Delta G (S = 1)$ obtained from the
MD results at 375~K, and various supersaturations.  Since $\Delta G (S
= 1)$ is supersaturation independent the values from all runs should overlap at all sizes.
However, our simulation data is only good enough for accurate abundance estimates below
a certain cluster size, which depends on the run properties.
The highest nucleation rates run of these (T375a) produced a large number of clusters over the entire plotted size range and
allows the most reliable reconstruction of $\Delta G_i (S)$.
The results from the other runs are only accurate at smaller sizes, where they overlap with (T375a).
Figure~\ref{fig:delgs-t375} also shows the surface energy $ \Delta
G_i(S = 1)$ divided by that of the CNT, i.e., $ \Delta G_i(S = 1) /(\eta
i^{2/3} kT)$.  In the figure, we also show the results of the SP model,
given by 
\begin{equation}
 \Delta G_i(S = 1) /(\eta i^{2/3} kT)=1+(\xi/\eta)
i^{-1/3} - (\xi/\eta) i^{-2/3}.
\end{equation}
The simulation results deviate from
the SP model at 375~K.  In Figure~\ref{fig:delgs-t375}, we can fit the
reconstructed $\Delta G_i (S = 1)$ with 
\begin{equation} \label{eq:kyoko}
\Delta G_i(S = 1)
/(\eta i^{2/3} kT) = 1+A i^{-1/3} - A i^{-2/3},
\end{equation} 
using a fitting parameter $A=0.9$ for small clusters.

\begin{figure}[]
\includegraphics[scale=0.5]{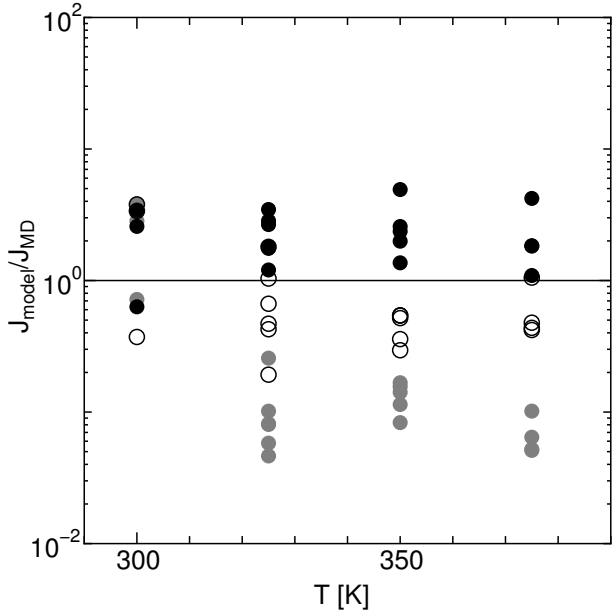}
\caption{Comparisons of nucleation rate from the MD simulations and several model predictions:
the SP model (grey filled circles), our new surface term fit (Eq. \eqref{eq:kyoko}) with $\alpha = 1$ (filled circles)
and using the $\alpha$ values measured in the MD simulations (open circles).}\label{fig:compare-j-new-alpha2}
\end{figure}

Figure~\ref{fig:compare-j-new-alpha2} shows the ratios between the
model and \eqref{eq:kyoko}
($A$=0.9, 1.0, 1.0 and 1.5 at 375, 350, 325, and $300\,$K,
respectively) and the MD simulations for two cases: one in which
$\alpha=1$ and the other in which $\alpha$ is set to be value obtained
directly from simulation.  In Figure~\ref{fig:compare-j-new-alpha2},
the predictions from the SP model are also shown for comparison.  We find
the new model agrees with the simulations within one order of
magnitude for all cases. At 375 K this is no surprise, since this data was used to
to determine the parameters of our fitting function for the surface term \eqref{eq:kyoko}.
The good agreement at the other temperatures is encouraging and might motivate
using \eqref{eq:kyoko} also to predict nucleation rates at different temperatures and
supersaturations. 

\section{Cluster Densities}\label{sec:densities}
It has been shown\citep{chapelar} that for spherical clusters, liquid-vapor interface densities are well-approximated by 
\begin{equation}\label{chapela}
\rho\left(r\right) = \frac{1}{2} \left[\rho_c + \rho_g -\left(\rho_c -\rho_g \right)\tanh \left(2 \frac{r-R}{d} \right) \right],
\end{equation} 
where $\rho_c$ is the number density within the cluster, $\rho_g$ the gas number density, $R$ the interface position, and $d$ its width. In each cluster's center-of-mass frame, we bin the spherical number density, using a bin size of $1.5 \textrm{\AA}.$ The number density profiles for clusters of the same size are used to make ensemble averages, to which equation \eqref{chapela} can be fit. This method of measuring internal cluster densities is robust only for clusters which are large enough to possess a constant density core. Clusters with $i<20-30$ are unlikely to have reached a shape well-describable by \eqref{chapela}, larger clusters' density profiles on the other hand are well-suited to this functional form. Density profile measurements are noisy, and so particularly large runs with many clusters in each size bin are necessary for the ensemble average to provide acceptable accuracy. For this reason we perform the density profile measurements in our largest simulation, T325f, and we do so at the end of the run. Figure \ref{fig:cluster_densities} plots $\rho_c$ against $R$ for clusters in T325f.  We observe an over-density for clusters between $4-5.5\textrm{\AA}$, however the clusters approach the bulk liquid values as they grow, although there seems to be a weak overdensity indication of $\sim 5\%$. This is in contrast to recent Lennard-Jones nucleation simulations\citep{mdlj4} which showed cluster densities significantly lower than the bulk liquid values. For post-critically sized clusters this was attributable to the increased cluster temperatures, due to the residual latent heat which had not been efficiently redistributed back into the gas. We surmise that the good agreement our internal cluster densities have with the bulk liquid values to be due to the fact that they are in thermal equilibrium with the surrounding gas.

In the Lennard-Jones case\citep{mdlj4}, the lowered densities for clusters with $i=i^*$ implied larger surface areas, and therefore larger-than-expected surface energies, resulting in an increased free energy cost to form a critical cluster, which lowered nucleation rates from model predictions. We suspect that nucleation rate predictions are more successful for SPC/E water than they are for Lennard-Jones because the assumption of small clusters possessing the bulk density for $i=i^*$ is more realistic for the case of SPC/E water. 

\section{Temperatures}\label{sec:temps}

\begin{figure}
\includegraphics[scale = 0.53]{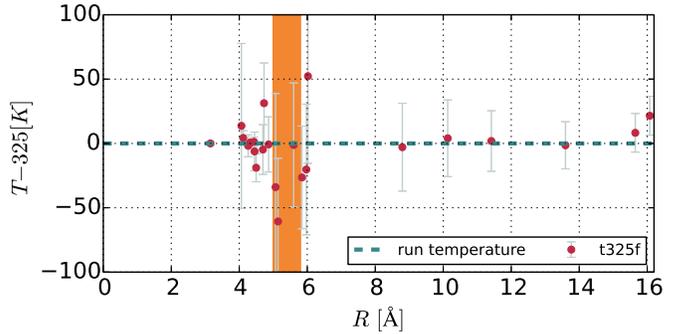}
\caption{Cluster temperatures for T325f. We observe cluster temperatures consistent with the run target temperature. There is no signal of residual latent heat for post-critical clusters, contrary to what has been observed in Lennard-Jones nucleation simulations\citep{mdlj4}. We suspect this may be due to the long-range nature of the SPC/E interaction potential, which enables efficient energy exchange between members of the cluster and the gas.}\label{fig:cluster_temps}
\end{figure}

\begin{figure}
\includegraphics[scale = 0.53]{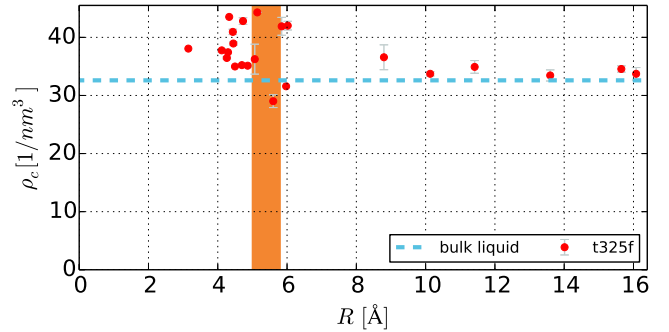}
\caption{Cluster densities for T325f, plotted against interface positions, both determined from density profile fits to \eqref{chapela}. We observe an over-density for clusters between $4-5.5\textrm{\AA}$, however the clusters approach the bulk values as they grow. We are unable to robustly implement the fitting procedure for very small clusters $<2.5\textrm{\AA}$ as they are not yet large enough to have converged to this shape, nor are they spherical. The orange region shows the expected critical cluster size, as estimated from the first nucleation theorem.}\label{fig:cluster_densities}
\end{figure}

We define the temperature of an ensemble of atoms from their mean kinetic energy
\begin{equation}
kT\equiv \frac{2}{3}\langle E_{\textrm{kinetic}} \rangle = \frac{1}{3N}\sum^N_{i=1} m v_i^2.
\end{equation}
Using full per-particle velocity information outputted at the end of the simulation, we are able to investigate the cluster size dependence of temperature. We find that sub-critical clusters are at the run average temperature, as observed in similar Lennard-Jones simulations\citep{mdlj4}. However, contrary to what has been observed in Lennard-Jones nucleation simulations, post-critical SPC/E water clusters possess temperatures consistent with the run average
 temperature. Figure \ref{fig:cluster_temps} plots the ensemble average (at each cluster size $i$) of their temperatures against the density profile interface midpoint $R$ (i.e., the cluster radius). 

The latent heat from condensation has been efficiently dissipated back into the gas, leaving the post-critical clusters in thermodynamic equilibrium with their surroundings. This finding is consistent with the post-critical clusters density profile measurements, which finds their densities at the expected bulk density. Had the clusters significant latent heat retention, their densities would be correspondingly lower. We conjecture that the efficient kinetic energy exchange is effected by the long-range Coulombic interactions - even molecules deep within a cluster may exchange energy and angular momentum with members of the gas - resulting in kinetic energy equipartition on shorter timescales than cluster growth rates. A molecule impinging on a cluster imparts heat onto into the cluster-system, yet the heat does not linger. This may have implications for non-isothermal nucleation models\citep{feder}, which include latent heat retention in the thermodynamic description of growing droplets. 

\section{Conclusions}

We have performed molecular dynamics simulations of SPC/E water, and significantly closed the nucleation-rate gap between simulation and experiment, measuring nucleation rates low as $\sim 10^{-19}\,\textrm{cm}^{-3}\textrm{s}^{-1}$. This is an hitherto unexplored saturation and temperature region for water nucleation experiments and water nucleation simulation. Nucleation rate results in this new regime will provide models with further testing comparison opportunities, to complement the already-existing lower nucleation rates from experiment, and higher nucleation rates from other simulations. We summarize our most significant contributions below.
\begin{itemize}
\item We introduce a new functional form, Eq. \eqref{eq:nucleation_rate} in order to implement the Yasuoka-Matsumoto nucleation rate measurement. This modified version smoothly captures the system's transition between the lag phase, relaxation phase, and onto the steady-state regime. 
\item As expected, the CNT over-estimates nucleation rates by a few orders of magnitude. The empirical CNT correction factor \eqref{eq:cnt_corr} \citep{correction_formula}, when using the Manka et al.\citep{water_experiment1} best-fit parameter values (calibrated in the low nucleation rate, low saturation regime) under-estimates our rates by a few orders of magnitude. When fitting their proposed correction function to our results, we find that the slopes are not steep enough. We conclude that this empirical and purely temperature-dependent correction factor to the CNT is not rich enough to reproduce the qualitative behavior we observe in our regime.
\item The MCNT nucleation model continues to over-predict nucleation rates, by factors of up to $10^4$. The SP model on the other hand, does somewhat better, under-predicting rates at worst by a factor of 24. Despite these failings, we note that these model predictions are significantly more accurate than the corresponding predictions for the Lennard-Jones fluid vapor-to-liquid nucleation\citep{mdlj2,mdlj3,mdlj4,mdlj5,mdlj6}. 
\item Performing a cluster growth rate measurement simulation under laboratory conditions (those found in Brus et al. (2008)\citep{theory_comparison10}) of $T=325\,$K and $S=2.5$, we measure a sticking probability of $\alpha = 0.15$. This suggests that in this regime, nucleation rate predictions from models should lowered by a factor of seven.
\item We find the cluster size $i\left(t\right)$ to be strongly cubic within the steady-state regime. The cluster growth rate is therefore proportional to the surface area, a result new to water nucleation. This is consistent with what has been found in Lennard-Jones nucleation simulations\citep{mdlj2,mdlj3}.
\item Unlike Lennard-Jones nucleation simulations, we find that post-critical clusters have temperatures consistent with the simulation average temperature: Growing clusters are in thermal equilibrium with their surroundings. Latent heat is not retained as the clusters grow, it is efficiently dissipated back into the gas. We suspect this efficiency is due to the long-range Coulombic interactions, not present in the Lennard-Jones case. This could have an impact on nucleation models which include non-isothermal processes into the thermophysical modeling of cluster properties\citep{feder}.

\item Post critical clusters have densities consistent with what is expected from the bulk liquid. There is a possible indication of an over-density for clusters around the critical size. This would imply a lower-than-expected surface area, which lowers the total surface energy, decreasing the free energy cost to form a critically-sized cluster and would result in higher-than-expected nucleation rates.  
\item The scaling relation $\ln S /(T/T_{\rm c}-1)^{1.7}$ is remarkably successful in reducing the 3-parameter $T\,$vs.$\,S\,$vs.$\,J$ surface into a 2-parameter curve. It accurately links nucleation rates from simulation and experiment from over 30 orders of magnitude in the nucleation rate range, and a temperature range of $180\,$K.
\end{itemize}

\bibliographystyle{apsrev4-1}

\bibliography{ms}
\begin{acknowledgments}
Computations were performed on Piz Daint at CSCS, and on the zBox4 at UZH. We thank the entire CSCS team. J.D. and R.A. are supported by the Swiss National Science Foundation. KKT is supported by JSPS KAKENHI Grant Number 2540054, 15K05015, and 15H05731. We thank the referees for their suggestions and comments.
\end{acknowledgments}

\end{document}